\documentstyle [12pt,epsf]{article}

\newcommand{\beq}{\begin{equation}}
\newcommand{\eeq}{\end{equation}}
\newcommand{\beqn}{\begin{eqnarray}}
\newcommand{\eeqn}{\end{eqnarray}}
\newcommand{\dd}{\partial}

\newcommand{\phib}{\bar{\phi}}

\newcommand{\etab}{\bar{\eta}}

\newcommand{\postscript}[2]
 {\setlength{\epsfxsize}{#2\hsize}
\setlength{\epsfysize}{#2\vsize}
  \centerline{\epsfbox{#1}}}

\begin{document}
\baselineskip 7.5 mm

\def\thefootnote{\fnsymbol{footnote}}
\baselineskip 7.5 mm

\begin{flushright}
\begin{tabular}{l}
UPR-671-T \\
hep-ph/9506386 \\
June, 1995
\end{tabular}
\end{flushright}

\vspace{20mm}

\begin{center}

{\Large \bf Tunneling in quantum field theory with spontaneous symmetry
breaking. }
\\

\vspace{20mm}

\setcounter{footnote}{0}

Alexander Kusenko\footnote{ email address: sasha@langacker.hep.upenn.edu}
\\
Department of Physics and Astronomy \\
University of Pennsylvania \\
Philadelphia, PA 19104-6396 \\

\vspace{30mm}

{\bf Abstract}
\end{center}

Tunneling in quantum field theory is well understood in the case of a
single scalar field.  However, in theories with spontaneous symmetry
breaking, one has to take into account the additional zero modes which
appear due to the Goldstone phenomenon.  We calculate the false vacuum
decay rate in this case.  The result may differ significantly from the
tunneling rates in the absence of symmetry breaking.

\vfill

\pagestyle{empty}

\pagebreak

\pagestyle{plain}
\pagenumbering{arabic}
\renewcommand{\thefootnote}{\arabic{footnote}}
\setcounter{footnote}{0}

\pagestyle{plain}

The effects of tunneling between different vacua in quantum field theory
play an important role in high-energy physics, cosmology and condensed
matter physics.  The rate of the false vacuum decay was calculated in
Refs. \cite{vko,c,cc} for a potential which depends on a single scalar
field.  However, most theories beyond the Standard Model, {\it e.\,g.},
the MSSM, grand unified theories, and strings, employ  scalar
potentials which depend on several scalar fields, and may have degenerate
minima related by some internal symmetry.  We will see that in this case
the results of Refs. \cite{vko,c,cc} do not apply because of the
complications that arise due to the additional zero-modes associated with
internal symmetries.  The purpose of this letter is to generalize the
semiclassical calculation of Refs. \cite{c,cc} to such cases.

If the potential $U(\phi)$ depends on a single scalar field $\phi(x)$, then
the corresponding path integral,

\beq
\int [d\phi] \  e^{-S[\phi]},
\label{path}
\eeq
is dominated by the field configuration
$\phib(x)$ called the ``bounce'' and can be evaluated using the saddle
point method  \cite{c,cc}.  Then the transition rate per unit
volume, in the semiclassical limit \cite{cc} is

\beq
\Gamma/{\sf V}=\frac{1}{h^2} (S[\phib])^2 e^{-S[\phib]/\hbar}
\left | \frac{\det'[-\dd_\mu^2+U''(\phib)]}{\det[-\dd_\mu^2+U''(0)]}
\right |^{-1/2}
\times (1+O(\hbar))
\label{one}
\eeq
where $S[\phib]$ is the Euclidean action of the bounce, and $\det'$ stands
for the determinant with the four zero eigenvalues omitted. These four zero
modes are related to the translational invariance in the four-dimensional
Euclidean space.  Integration over each of them yields a factor of
$\sqrt{S[\phib]}$ in equation (\ref{one}).

Let us now consider a quantum field theory with a scalar potential
$U(\phi_1, ... , \phi_n)$ which has a local minimum
at $\phi_1=\phi_2=...=\phi_n=0$, $U(0,...,0)=0$ as well as at least one
additional (local, or global) minimum at $\phi_i=\phi_i^e, \ i=1,2,...,n$;
$U(\phi_1^e,...,\phi_n^e)<0$.

The bounce, $\phib(x)=(\phib_1(x), ... ,\phib_n(x) )$, is the stationary
point of the action, so that  $\delta S[\phib]=0$.  It is found as an
O(4)-symmetric \cite{c_comm} solution $\phib(r)$, $r=\sqrt{x_\mu x^\mu}$,
of the corresponding Euler-Lagrange equations\footnote{In most cases, one
cannot solve this system of non-linear equations analytically.  An
effective method for finding the bounce was proposed in Ref. \cite{ak}.}:

\beq
   \Delta \phib_i(r)= \frac{\dd}{\dd \phib_i} U(\phib_1, ... , \phib_n)
\label{bounce_eq}
\eeq
with the following boundary conditions which ensure the finiteness of
$S[\phib]$:

\beq
\left \{ \begin{array}{l}
    (d/dr) \phib_i(r)|_{r=0}=0 \\  \\
   \phib_i(\infty)=0
        \end{array} \right.
\label{bc}
\eeq

\begin{figure}
\postscript{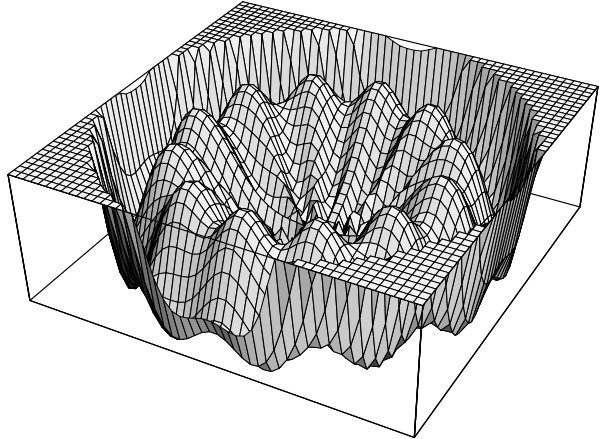}{0.5}
\caption{The potential {$U(\phi_1,\phi_2)$} with several degenerate minima
related by a {$Z_m$} symmetry}
\label{fig1}
\end{figure}

If the potential $U(\phi)$ has only one local minimum characterized by a
negative value of the potential, equation (\ref{one}) gives the correct
answer for the rate of the false vacuum decay.  Let us now consider the
potential $U(\phi_1,\phi_2)$ which has $m$ degenerate local minima related
by the $Z_{_m}$ symmetry as shown in Fig. 1.  The equation
(\ref{bounce_eq}) will have $m$ independent solutions satisfying the boundary
conditions (\ref{bc}), each bounce pointing in the direction of a given
minimum in group space.  Tunneling to different vacua are independent
events, and therefore the total probability of the false vacuum decay
is the sum of the contributions from each individual bounce.  The naive
application of the formula (\ref{one}) might lead one to conclude that as
the number of local minima $m$ increases, the lifetime of the false vacuum
would approach zero.  If one then considers the case of a broken $U(1)$
symmetry (Fig. 2) as the limit $m\rightarrow \infty$, one might argue that
the false vacuum would decay infinitely fast because there are infinitely
many channels, each having a finite probability determined by formula
(\ref{one}). This conclusion is wrong because, in the case of a continuous
symmetry, the Goldstone modes give rise to some additional zero modes in
the determinant in  equation (\ref{one}), so that the right-hand side of
(\ref{one}) is, in fact, divergent.

Clearly, formula (\ref{one}) cannot be used to describe the tunneling
probability in theories with spontaneously broken continuous symmetries.
One must not include the Goldstone zero modes in the determinant,
but deal with them instead in a similar manner to the way one usually
treats the zero eigenvalues associated with translational invariance.
The latter is the basic idea of our calculation.

We begin by considering the potential $U(\phi_1,\phi_2) \equiv
U(\phi_1^2+\phi_2^2)$ which has a $U(1)$ symmetry (Fig. 2).
We suppose that in addition to the local minimum at $\phi=0$,
$U(\phi)$ has a $U(1)$ orbit of degenerate global minima at some non-zero
value of $\phi$.  It is convenient to redefine the fields in terms of the
radial and the angular components:

\beq
(\phi_1(x),\phi_2(x))=(\eta(x) Re(e^{i\theta(x)}),\eta(x)
Im(e^{i\theta(x)}))
\eeq

Then $U(\phi)=U(\eta), (\dd/\dd \theta)U=0$, and every solution of the
equation (\ref{bounce_eq}) can be written in the form

\beq
\phib(x)=e^{i \theta} \etab(x)
\eeq
where $\theta$ is a constant independent of $x$ and $\etab(x)$ is the usual
bounce, the same as in the one-component case.

\begin{figure}
\postscript{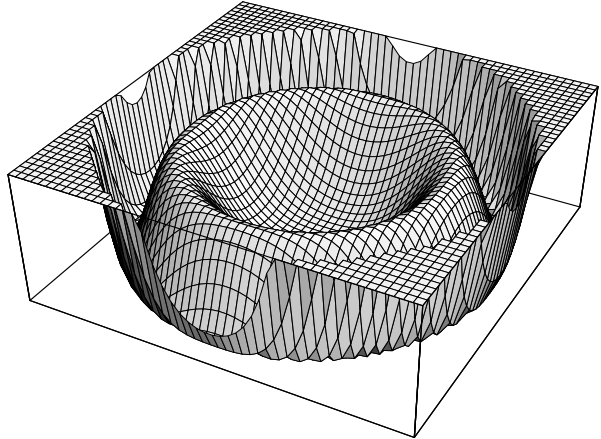}{0.5}
\caption{The potential {$U(\phi_1,\phi_2)\equiv U(\phi_1^2+\phi_2^2)$} with
a U(1) orbit of degenerate minima.}
\label{fig2}
\end{figure}

The idea of the method, used in Refs. \cite{c,cc} to calculate the
transition probability, was to evaluate the path integral using the saddle
point method.  The bounce $\phib(x)$ is a stationary point of the action
($\delta S[\phib]=0$), but it is not a minimum because the second variation
operator

\beq
\frac{\delta S}{\delta \phi_i \delta \phi_j}= -\delta_{ij}\Delta+
\frac{\dd^2U(\phi)}{\dd \phi_i \dd \phi_j}
\label{d2S}
\eeq
has one negative eigenvalue.  In addition, this operator has four zero
eigenvalues due to translational invariance in the coordinate space.  The
corresponding normalized eigenfunctions are \cite{c,cc}

\beq
\phi_\mu=(S[\phib])^{-1/2} \dd_\mu \phib(x), \ \ \mu=1,...,4
\label{0x}
\eeq

In the vicinity of the bounce, one can expand the field $\phi(x)$ in the
(complete orthonormal) basis of the eigenvectors of the operator
(\ref{d2S}):

\beq
\phi_k(x)=\phib_k(x)+\sum_{i} c_k^{(i)} \phi_k^{(i)}
\label{expand}
\eeq

Then the path integral measure, $[d\phi_k]$, is defined as

\beq
[d\phi_k]=\prod_i (2\pi \hbar)^{^{-\frac{1}{2}}} dc_k^{(i)}
\label{path_measure}
\eeq

The path integration with respect to the Fourier coefficients $c_k^{(i)}$,
which correspond to positive modes, is Gaussian and results in the
determinant of the type (\ref{one}) where the zero modes are not included.
The integration over zero modes, however, is less straightforward.  We
would like to treat the Goldstone modes on the same footing as the Poincare
zero modes (\ref{0x}).  The first step is to find the orthonormalized basis
in the subspace of eigenvectors that correspond to zero eigenvalues.
By explicit differentiation of equation (\ref{bounce_eq}) with respect to
$\theta$ we find that, as long as $U(\phi) \equiv U(\eta)$ is independent of
$\theta$,

\beq
\phi_\theta =\left (\int \etab^2(x) d^4x \right )^{-1/2} \dd_\theta \phib(x)
\label{0t}
\eeq
is an eigenvector of (\ref{d2S}) corresponding to a zero eigenvalue.  The
normalization comes from
$\int |\dd_\theta \phib(x)|^2 d^4 x=\int  \etab^2(x) d^4x$.  It is easy to
see that the eigenvector (\ref{0t}) is orthogonal to each of the
eigenvectors (\ref{0x}).  Therefore, the set of functions (\ref{0x}) and
(\ref{0t}) combined form a complete orthonormal basis in the subspace of
zero modes of the operator (\ref{d2S}).

Let $c_\theta$ be the Fourier coefficient in (\ref{expand}) corresponding
to the Goldstone mode. Then the change in $\phi$ corresponding to some small
change in $c_\theta$ is $d\phi= \phi_\theta dc_\theta$.  On the other hand,
$d\phi = \dd_\theta \phib \ d\theta$.  Comparing these relations to each
other, and using the normalization from equation (\ref{0t}), one gets

\beq
(2\pi \hbar)^{-1/2} dc_\theta= (\int \etab^2(x) d^4x  /
2\pi \hbar)^{1/2} d\theta
\label{dtheta}
\eeq

Therefore, integration with respect to $c_\theta$ in the path integral
(\ref{path}) is equivalent to the integration over the angle $\theta$ in
group space, up to the factor of $\sqrt{\int \etab^2}$.  The field
configuration that comprises $n$ widely separated bounces is also a saddle
point of the path integral with the action $n S[\phib]$. We have to sum
over all such configurations and integrate over the positions of the
centers of the bounces.  This is done in full analogy to the corresponding
calculation in \cite{c,cc}.

The result is

\beq
\Gamma/{\sf V}= 2\pi \left [\int \etab^2(x) d^4x \right ]^{1/2}
\left(\frac{S[\etab]}{2\pi \hbar} \right)^2
e^{-S[\etab]/\hbar}
\left | \frac{\det''[-\dd_\mu^2+U''(\etab)]}{\det[-\dd_\mu^2+U''(0)]}
\right |^{-1/2}
\times (1+O(\hbar))
\label{u1}
\eeq
where $\det''$ stands for the determinant from which all zero eigenvalues,
including the Goldstone zero mode, are excluded.

The result (\ref{u1}) can be easily generalized to the case of a larger
internal symmetry group.  Suppose the true vacuum is characterized by the
vev that breaks some symmetry group $G$ down to $G'$, so that the scalar
potential has $N=rank(G/G')$ Goldstone modes.  The false vacuum decay rate
is then

\beq
\Gamma/{\sf V}= C(G/G') \left [\int \etab^2(x) d^4x \right ]^{N/2}
\left(\frac{S[\etab]}{2\pi \hbar} \right)^2
e^{-S[\etab]/\hbar}
\left | \frac{\det''[-\dd_\mu^2+U''(\etab)]}{\det[-\dd_\mu^2+U''(0)]}
\right |^{-1/2}
\times (1+O(\hbar))
\label{G}
\eeq

Here $C(G/G')$ is a dimensionless factor that depends on the quotient group
$G/G'$.  The pre-exponential factor in square brackets has dimension
$N$ in length units, and the determinant in the numerator, with $4+N$ zero
modes omitted, has dimension $2(4+N)$ in excess of that of the denominator.
The power $(-1/2)$ reduces the latter to $-(4+N)$, so that the total
dimension of the right-hand side is minus four, as it should be.

Some comments are in order. We note that the result (\ref{G}) can differ
significantly from that
obtained by using the formula (\ref{one}) and ignoring the symmetries of
the potential.  For example, let us consider the case of
$G=SU(5) \rightarrow G'=SU(3)\times SU(2) \times U(1)$. There are $N=12$
independent Goldstone modes in this case.  In the thin wall approximation,
one can express $\int \etab^2$ in terms of $S[\etab]$ times some dimensionful
quantity.  Then it can be shown that the answer obtained by applying
formula (\ref{one}) and ignoring the Goldstone modes would be off by a
factor of $(S[\phib])^6$.  For $S[\phib] \sim 10-100$, the difference
would be as great as six to twelve orders of magnitude.

We also note that the increase in the number of degenerate minima $m$ (Fig.
1), does not necessarily imply an increase in $\Gamma$.  In fact,
to accommodate $m$ degenerate minima, $m \rightarrow \infty$, separated by
the fixed height barriers, one has to allow the curvature (in the
$\theta$-direction) at the bottom of each minimum to grow as $\sim m^2$.
This corresponds to $\sim (1/m) $ behavior of the determinant factor in
equations (\ref{u1}) or (\ref{G}).  Now the total decay rate is $m$ times
the contribution of each minimum, $\Gamma \sim \Gamma_{_0} m (1/m) \rightarrow
const$.  This is the resolution of the paradox posed in the beginning.

It is straightforward to generalize formula (\ref{G}) to the case of
quantum field theory at finite temperature \cite{linde}.

We have assumed that $G' \subset G$, which is the case for
many applications in particle physics.  However,  in some cases
the true vacuum may, at some temperature, have a larger symmetry
than the false vacuum (see, {\it e.\,g.}, Ref. \cite{lp}), so that
$G \subset G'$.  Finally, the true vacuum may be characterized by
the symmetry group, $G'$, completely unrelated to that of the false vacuum,
$G$.  If this is the case, then in general one has to modify the expression
(\ref{G}) to take into account the particular features of the potential.

To summarize, we have calculated the tunneling probability in quantum field
theory with spontaneously broken local or global symmetries.

The author would like to thank P. Langacker for many stimulating
discussions and helpful comments.  We are also grateful to G. Segre and P.
Steinhardt for useful conversations.  This work was supported by the
U.~S. Department of Energy Contract No. DE-AC02-76-ERO-3071.

\end{document}